\documentclass[letterpaper, 10 pt, conference]{ieeeconf}  
\usepackage{graphicx,graphics}
\usepackage{amsmath,amsfonts,mathrsfs,amssymb,dsfont}
\usepackage{booktabs}
\usepackage{cite}
\usepackage{bbm}
\usepackage{color}
\usepackage{arydshln}

\newcommand{\et}{\hfill \blacksquare}
\newtheorem{theorem}{\bf{Theorem}}

\IEEEoverridecommandlockouts                              

\title{\LARGE \bf
Linear piecewise-deterministic Markov processes with families of random discrete events}

\author{Mohammad Soltani$^{1}$, Abhyudai Singh$^{2}$
\thanks{$^{1}$M. Soltani is with Department of Electrical and Computer Engineering, University of Delaware, Newark, DE USA 19716.
{\tt\small msoltani@udel.edu}}%
\thanks{$^{2}$A. Singh is with the Department of Electrical and Computer Engineering, Biomedical Engineering, Mathematical Sciences, Center for Bioinformatics and Computational Biology, University of Delaware, Newark, DE USA 19716.
{\tt\small absingh@udel.edu}}}

\begin{document}

\maketitle
\thispagestyle{empty}
\pagestyle{empty}

\begin{abstract}
We consider a class of piecewise-deterministic Markov processes where the state evolves according to a linear dynamical system. This continuous time evolution is interspersed by discrete events that occur at random times and change (reset) the state based on a linear affine map. In particular, we consider two families of discrete events, with the first family of resets occurring at exponentially-distributed times. The second family of resets is generally-distributed, in the sense that, the time intervals between events are independent and identically distributed random variables that follow an arbitrary continuous positively-valued probability density function. For this class of stochastic systems, we provide explicit conditions that lead to finite stationary moments, and the corresponding exact closed-form moment formulas. These results are illustrated on an example drawn from systems biology, where a protein is expressed in bursts at exponentially-distributed time intervals, decays within the cell-cycle, and is randomly divided among daughter cells when generally-distributed cell-division events occur. Our analysis leads to novel results for the mean and noise levels in protein copy numbers, and we decompose the noise levels into components arising from stochastic expression, random cell-cycle times, and partitioning. Interestingly, these individual noise contributions behave differently as cell division times become more random. In summary, the paper expands the class of stochastic hybrid systems for which statistical moments can be derived exactly without any approximations, and these results have applications for studying random phenomena in diverse areas.
\end{abstract}
\section{INTRODUCTION}
 
Hybrid systems that couple continuous dynamics with discrete random events are increasingly being used for studying stochasticity across disciplines 
\cite{HespanhaMar04,BohacekHespanhaLeeObraczkaJun03,jia06,ahs13,hes14,bur11,vgl06,lih11,sma12,adl12,hsi04}. These systems fall into the category of piecewise-deterministic Markov processes (PDMPs) when the continuous dynamics is deterministic, and often modeled through differential equations \cite{dav84,cos90,dav93,cod08,flj92,fgd00,cfm08}. Given their important and ubiquitous usage, a wide range of analysis tools have been developed for PDMPs \cite{kam11,sih10a,abg14,ddd16,ddz15}.  While computing statistical moments is straightforward for linear stochastic systems, any form of nonlinearities within PDMPs leads to the non-trivial problem of unclosed moment dynamics: time evolution of lower-order moments depends on higher-order moments. Typically, closure techniques \cite{kon12,smk13,gri12,sih10, svs15,gsl17}, or constraints imposed by positive semidefiniteness of moment matrices \cite{gvl17, lgs16} are  exploited to solve moments in such cases.

Our prior work has identified classes of stochastic hybrid systems, where time evolution of moments can be obtained exactly without any approximations \cite{sos16c,sos17,sos15}. While much of this work involves a linear continuous-time dynamical system with a single family of random discrete events, here we expand this work to consider families of discrete events, and in particular events with memory that occur at non-exponential time intervals. For these systems, we provide exact analytical solutions for the first- and second-order moments, and our approach can be generalized to derive any arbitrary high-order moment (skewness, kurtosis, etc.). 

As an interesting application, we use PDMPs to model one of the most fundamental processes in cell biology: production and decay of protein molecules in an individual living cell. We propose a model that incorporates noise in protein copy numbers from three different mechanisms: random timing of cell-division events that reset the protein levels by approximately half; random partitioning of protein molecules between two daughter cells at the time of division, and randomness in protein production that has been well documented across organisms. Our analysis provides for the first time, an exact formula for the mean and noise in protein levels, and we discuss novel insights obtained from these results. For example, in some parameter regimes making cell-division times more random can reduce the extent of random fluctuations in protein counts. We start by giving a mathematical description for the PDMPs under investigation, go on to provide results on statistical moments, and finally, discuss their application on the biological example.

\section{Model formulation}
The class of PDMPs under consideration have the following ingredients: 
\begin{enumerate}
	\item \textbf{Continuous dynamics}:
The states of the system $\bold x \in \mathbb{R}^{n \times 1}$ are governed by time-invariant ordinary differential equations (ODEs)
\begin{equation}
\dot{\bold x }(t)= \hat{a}+A \bold x, \label{dynamics0000}
\end{equation}
where vector $a \in \mathbb{R}^{n\times 1}$ and matrix $A \in \mathbb{R}^{n\times n}$ are constant. While exact moment computations can be easily extended to linear 
stochastic differential equations, we prefer to work with ODEs for the sake of simplicity. 

\item \textbf{Exponentially-distributed resets}:
The first family of resets occur at exponentially-distributed time intervals, i.e., Poisson arrival of events. Let the mean time interval in between these resets be denoted by $1/h_1$ where $h_1$ is a constant. Then, the probability of an event occurring in the next infinitesimal time interval $(t,t+dt]$ is $h_1 dt$. Whenever these events occur the state is reset based on a linear affine map
\begin{equation}
\bold x\mapsto J_1\bold x + R_1, \label{reset}
\end{equation}
where $J_1\in \mathbb{R}^{n\times n}$ and $R_1 \in \mathbb{R}^{ n \times 1}$ are a constant matrix and vector, respectively. 

\item \textbf{Generally-distributed resets}:
The second family of resets occur in non-exponentially distributed time intervals. Events occur at times $\boldsymbol{t}_s, \ s\in \{1,2,\ldots\}$, and the time intervals
\begin{align}
\boldsymbol T_s  \equiv   \boldsymbol t_s - \boldsymbol  t_{s-1}\label{pdf of T}
\end{align}
are independent and identically distributed (iid) random variables that follow an arbitrary continuous positively-valued probability density function $f$. Whenever the events occur, the state is reset as
\begin{equation}
\bold x\mapsto \bold x_{2+}. \label{reset2}
\end{equation}
We allow $\bold x_{2+}$ to be a random variable, whose average value is related to its value just before the event as 
\begin{align}
&\langle \bold{x}_{2+}\rangle=  J_2\bold{x}+ R_2.\label{conditional x0}
\end{align}
Here $\langle . \rangle $ denotes the expected value operator,  $J_2\in \mathbb{R}^{n\times n}$ and $R_2 \in \mathbb{R}^{ n \times 1}$ are a constant matrix and vector, respectively.  Furthermore, the covariance matrix of $\bold x_{2+}$ is defined by
\begin{equation} 
\begin{aligned}
\langle \bold{x}_{2+} \bold{x}_{2+}^\top \rangle -  \langle \bold{x}_{2+}\rangle\langle & \bold{x}_{2+}\rangle^\top=Q_2 \bold{x}\bold{x}^\top Q_2^\top \\ &	\label{conditional x20}
+ B_2  \bold{x}C_2^\top+ C_2 \bold{x}^\top B_2^\top+ D_2.
\end{aligned}
\end{equation}
Here $ Q_2 \in \mathbb{R}^{n \times n}$, $ B_2 \in \mathbb{R}^{n \times n}$, $ C_2 \in \mathbb{R}^{n \times 1}$ are constant matrices. Moreover $ D_2 \in \mathbb{R}^{ n \times n}$ is a constant symmetric matrix. The matrices $Q_2$, $B_2$, $C_2$, and $D_2$ can be used to incorporate constant or state-dependent noise in $\bold  x$ at the time of the reset \cite{sva15,ss16,vss16}. 
\end{enumerate}

Having mathematically defined the system, we next model generally-distributed resets via a timer, and then provide our main results on the statistical moments of ${\bold x}(t)$.

\begin{figure}[!b]
\centering
{\includegraphics[width=0.7\columnwidth]{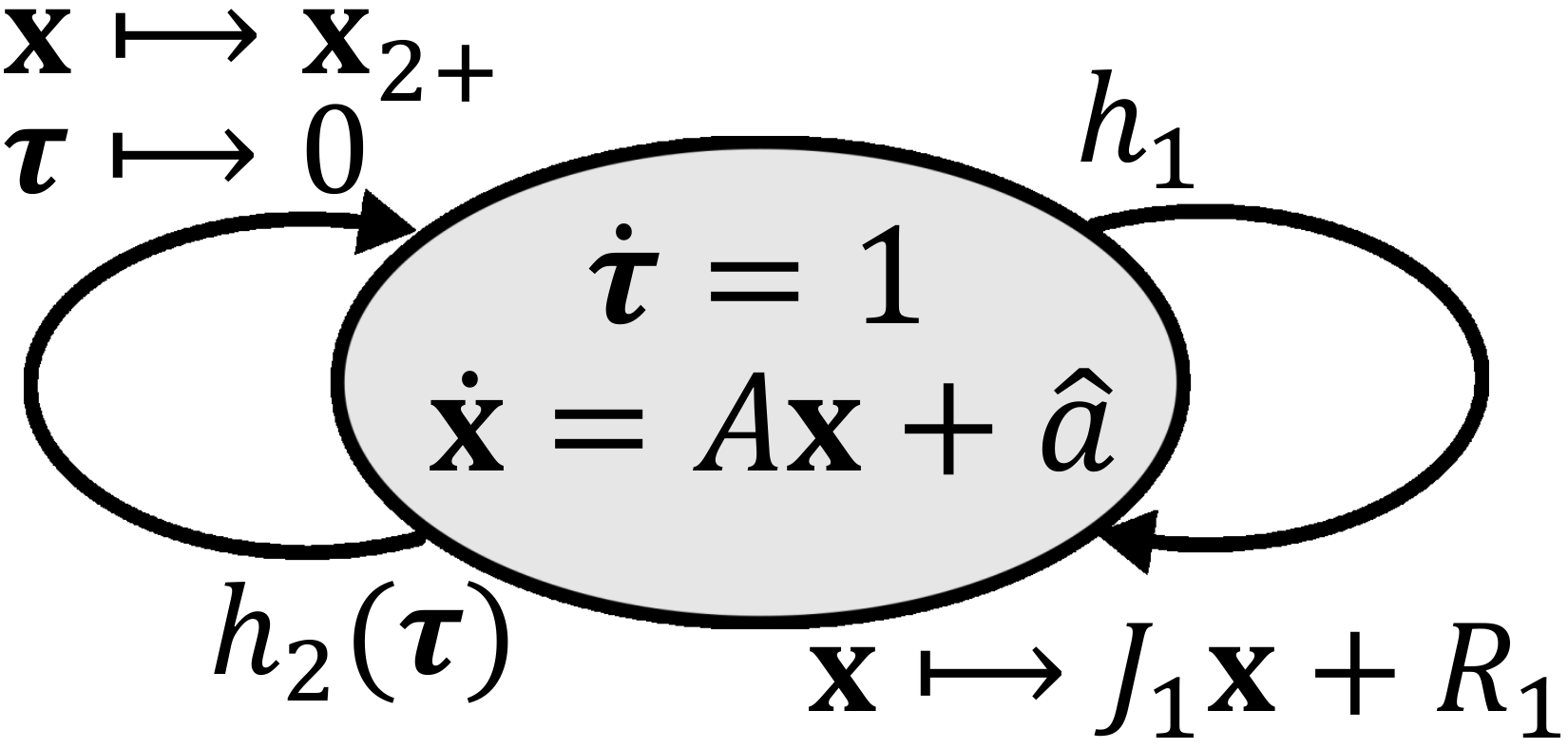}}
\caption{{\bf Schematic of a PDMP with linear continuous dynamics and two families of random resets}. The two families of resets occurs at random times and change the state based on the linear affine maps \eqref{reset} and \eqref{reset2}.  For the first reset, events happen at exponentially-distributed time intervals, while for the second reset, time intervals are generally-distributed and drawn from an arbitrary probability distribution function. The timing of the latter family of resets is regulated by a timer $\boldsymbol \tau$ that measures the time since the last event, and the next
event occurs at a hazard rate $h_2( \boldsymbol \tau )$. In between events, the state evolve according a linear time-invariant dynamical system.}
\label{figmodel}
\end{figure}

\section{Deriving Statistical moments}
We recast the above PDMP as a state-driven process by introducing a timer $\boldsymbol \tau$ that measures the time since the last generally-distributed event (Fig. 1). The timer increases with time in between the events 
\begin{equation}
\dot{\boldsymbol \tau} =1,
\end{equation}
and resets to zero whenever a new generally-distributed event occurs
\begin{equation}
\boldsymbol \tau \mapsto 0.
\end{equation}
With this definition, one can define a timer-dependent hazard rate that ensures the time interval $\boldsymbol T_s$ between two successive generally-distributed events follows
an arbitrary given probability distribution function \cite{Ross20109,ehp00}. In particular, let the probability of event occurrence in the next infinitesimal time $(t,t+dt]$ be $h_2(\boldsymbol \tau)dt$, where
\begin{equation} \label{hr}
h_2( \tau ) \equiv   \frac{f(\tau)}{1-\int_{y=0}^{\tau}f(y)dy}.
\end{equation}
Then, the time interval $\boldsymbol T_s$ follows the probability density function $f$ , and $\boldsymbol \tau$'s  probability density function is given by
\begin{equation}
p(\tau) = \frac{1}{\langle \boldsymbol T_s \rangle} {\rm e}^{-\int_0^{\tau} h(y) d y} \label{prob. tau}
\end{equation}
where $\langle \boldsymbol T_s \rangle$ is the average time between generally-distributed events (Appendix A). As expected, the function $h_2$ becomes timer independent (constant) for an exponential distribution $f$, but is typically nonlinear for most distributions. 

\subsection{Steady-state mean level}
To present the steady-state mean of $\bold x$, we define the following matrix and vector to simplify notation
\begin{equation}
A_{\overline{ x }} \equiv A+h_1 (J_1 -I_n)), \ \ \hat{a}_{\overline{ x }} \equiv \hat{a}+R_1,
\end{equation}
where $I_n$ is a $n$-dimensional identity matrix. 
\begin{theorem}
Consider the PDMP given by the linear continuous dynamics \eqref{dynamics0000}, and the two families of stochastic events defined in \eqref{reset}-\eqref{conditional x20}. For this system, the steady-state mean is finite, if and only if, all the eigenvalues of 
$ J_2  \left\langle  {\rm e}^{ A_{\overline{ x }}  \boldsymbol T_s } \right \rangle
$ are inside the unit circle, and in this case 
	\begin{equation}\label{mean level01}
	\begin{aligned}
 \overline{\langle \bold{x} \rangle} \equiv \lim_{t \to \infty} \langle \bold{x(t)} \rangle= &  \left\langle 	{\rm e}^{ A_{\overline{ x }}\boldsymbol \tau }  \right\rangle  \left(I_n -\left\langle  J_2 {\rm e}^{  A_{\overline{ x }}  \boldsymbol T_s } \right \rangle  \right)^{-1}  \times \\ &
\left( R_2  +  J_2  \left \langle  {\rm e}^{ A_{\overline{ x }} \boldsymbol T_s } \int_0^{\boldsymbol T_s } {\rm e}^{- A_{\overline{ x }} r} \hat{a}_{\overline{ x }} dr  \right \rangle \right)\\ & +  \left\langle     {\rm e}^{  A_{\overline{ x }}  \boldsymbol \tau }  
 \int_0^{\boldsymbol \tau}  {\rm e}^{-  A_{\overline{ x }} r}  \hat{a}_{\overline{ x }} dr   \right\rangle ,
	\end{aligned}
	\end{equation} 
where $\overline{\langle . \rangle}$ denotes the expected value at steady-state.
\end{theorem}$\et$\\
\noindent The sketch of the proof is the following: We start by finding the moment dynamics of the system in between generally distributed events.
Next, we solve moment dynamics for initial condition corresponding to $\boldsymbol \tau =0$ (value of the states right after a generally distributed reset). Finally, taking mean over all $\boldsymbol \tau $ results in the steady-state mean of $\bold x$. For a detailed proof please see Appendix B. Note that in this theorem we use the notation $\boldsymbol T_s$ when we take expected value with respect to $\boldsymbol T_s$ (e.g. $\left\langle  {\rm e}^{ A_{\overline{ x }}  \boldsymbol T_s}  \right \rangle $) and we use $\boldsymbol \tau$ when we take expected value with respect to $\boldsymbol \tau$ (e.g. $\left\langle  {\rm e}^{ A_{\overline{ x }}  \boldsymbol \tau}  \right \rangle $).

\subsection{Steady-state noise level}
To compute the second-order moments of $\bold x$, our first step is to transform the time evolution of $\bold x  \bold x^\top$ to a similar form as in \eqref{dynamics0000}
\begin{equation} \label{non vector}
\frac{d\left( \bold x  \bold x^\top\right)}{dt}=
\frac{d  \bold x}{dt}\bold x^\top+ \bold x \frac{d  \bold x^\top}{dt}= 
A\bold x  \bold x^\top+ \bold x  \bold x^\top A^\top+ \hat{a}  \bold x^\top+  \bold x \hat{a}^\top.
\end{equation}
We use vector representation of this equation by putting all the columns of the matrix $\bold x  \bold x^\top $ into one vector. Vectorization is a linear transformation which converts a matrix into a column vector. Using vectorization, \eqref{non vector} can be rewritten as 
\begin{equation}
\frac{d{\rm vec}\left(\bold x  \bold x^\top\right)}{dt} =  (I_n \otimes A + A\otimes I_n){\rm vec}\left({\bold x  \bold x}^\top\right) +( I_n \otimes \hat{a} + \hat{a}  \otimes I_n)\bold x ,
\end{equation}
where '${\rm vec}()$' stands for vector representation of a matrix and $\otimes$ denotes the Kronecker product. Since $\bold x  \bold x^\top\in \mathbb{R}^{n\times n}$ then ${\rm vec}\left(\bold x  \bold x^\top\right)\in \mathbb{R}^{n^2 \times 1}$. Moreover note that ${\rm vec}(\bold x)={\rm vec}(\bold x^\top)=\bold x$. Further in deriving this equation we used the fact that for three matrices $M_1$, $M_2$, and $M_3$ 
\begin{equation}
{\rm vec}(M_1 M_2 M_3) = (M_3^\top \otimes M_1){\rm vec}(M_2)
\label{kronecker}
\end{equation}
 \cite{mao13}. In the next step, we define a new vector $\boldsymbol \mu\equiv \left[\bold x^\top \ \ \ {\rm vec}  \left( \bold x  \bold x^\top \right)^\top\right]^\top\in \mathbb{R}^{(n+n^2) \times 1}$, whose dynamics in between events is governed via
\begin{equation}
\frac{d\boldsymbol \mu}{dt} =\hat{a}_\mu + A_\mu  \boldsymbol \mu, \label{mu dynamics}
\end{equation}
where 
\begin{equation}\small 
\begin{aligned}
& 
A_\mu\equiv & \left[\begin{array}{c;{2pt/2pt}c}
A & 0\\ \hdashline[2pt/2pt] I_n \otimes \hat{a} + \hat{a}  \otimes I_n & I_n \otimes A + A\otimes I_n 
\end{array}\right],  
a_\mu \equiv \left[\begin{array}{c}
\hat{a}\\ \hdashline[2pt/2pt] 0
\end{array}\right].
\end{aligned}
\end{equation}

Any time that an exponentially distributed event occurs, vector $\boldsymbol \mu$ resets as
\begin{equation}
\boldsymbol \mu \mapsto J_{\mu 1} \boldsymbol \mu+ R_{\mu 1}. \label{reset mu}
\end{equation}
where 
\begin{equation}\small
\begin{aligned}
& J_{\mu1} \equiv    \left[\begin{array}{c;{2pt/2pt}c}
J_1  & 0\\ \hdashline[2pt/2pt] 0 &  J_1\otimes J_1
\end{array}\right], \ \ \ 
R_{\mu 1}  \equiv  \left[\begin{array}{c}
R_1 \\ \hdashline[2pt/2pt]  {\rm vec }(R_1 R_1^\top ) 
\end{array}\right]. \label{Jmu}
\end{aligned}
\end{equation}
Moreover, any time that a generally-distirbuted event occurs, $\boldsymbol \mu$ resets as 
\begin{equation}
\boldsymbol \mu \mapsto \boldsymbol \mu_{2+}. \label{reset mu2}
\end{equation}
Based on \eqref{conditional x20}
\begin{equation}
\begin{aligned}
\langle \bold{x}_{2+} \bold{x}^\top_{2+} \rangle = & \langle \bold{x}_{2+}\rangle\langle \bold{x}_{2+} \rangle^\top+Q_2 \bold{x}\bold{x}^\top Q_2^\top\\ &	
+ B_2   \bold{x}C_2^\top+ C_2 \bold{x}^\top  B_2^\top+ D_2.
\end{aligned}
\end{equation}
Further from \eqref{conditional x0}, $\langle \bold{x}_{2+} \rangle\langle \bold{x}_{2+} \rangle^\top  $ can be written as
\begin{equation}\begin{aligned}
\langle \bold{x}_{2+} \rangle\langle \bold{x} _{2+} \rangle^\top   =  J_2  \bold{x}\bold{x}^\top J_2^\top
+ J_2   \bold{x} R_2^\top+ R_2 \bold{x}^\top  J_2^\top+ R_2R_2^\top.
\end{aligned}
\end{equation}
By combining these two equations and using \eqref{kronecker}, $\langle \boldsymbol\mu_{2+} \rangle $ is given by 
\begin{equation}
\begin{aligned}
\langle \boldsymbol \mu_{2+} \rangle =   J_{\mu 2} \boldsymbol \mu+ R_{\mu 2} , \label{reset 2}
\end{aligned}
\end{equation}
where 
\begin{equation}\small
\begin{aligned}
& J_{\mu 2} \equiv    \left[\begin{array}{c;{2pt/2pt}c}
J_2  & 0\\ \hdashline[2pt/2pt] \begin{array}{c} B_2 \otimes C_2  +J_2 \otimes R_2 \\  +C_2 \otimes B_2 +R_2 \otimes J_2 \end{array}  &  J_2 \otimes J_2 +Q_2 \otimes Q_2  
\end{array}\right], \\ &
R_{\mu 2}  \equiv  \left[\begin{array}{c}
R_2 \\ \hdashline[2pt/2pt]  {\rm vec }(D_2 +R_2 R_2^\top ) 
\end{array}\right]. \label{Jmu}
\end{aligned}
\end{equation}

Deterministic dynamics \eqref{mu dynamics}, and stochastic resets \eqref{reset mu} and \eqref{reset 2} are similar to  \eqref{dynamics0000}, \eqref{reset} and \eqref{conditional x0}. Hence with a similar analysis as in Theorem 1, the following theorem provides the necessary and sufficient conditions for having finite second-order moments of $\bold x$. As done prior to Theorem 1, we define
\begin{equation}
A_{\overline{ \mu  }} \equiv  A_\mu +h_1 (J_{\mu 1} -I_{n^2+n}), \ \ \hat{a}_{\overline{ \mu}}\equiv \hat{a}+R_{\mu 1 }, \label{mu bar}
\end{equation}
where $I_{n^2+n}$ is a $n^2+n$ dimensional identity matrix.
\begin{theorem}
Suppose that the states of the system given by \eqref{dynamics0000}-\eqref{conditional x20} satisfies Theorem 1. Then $\overline{ \langle \bold x  \bold x^\top\rangle}$ is finite if and only if all the eigenvalues of the matrix $(J_2\otimes  J_2 +Q_2\otimes Q_2) \left\langle  {\rm e}^{ A_\mu \boldsymbol T_s} \otimes {\rm e}^{ A_\mu  \boldsymbol T_s} \right \rangle $ are inside the unit circle. In this limit
	\begin{equation}\label{mean level}
	\begin{aligned}
 \overline{\langle \boldsymbol \mu \rangle}\equiv & \lim_{t \to \infty} \langle \boldsymbol{\mu} \rangle=  \left\langle 	{\rm e}^{ A_{\overline{\mu }}\boldsymbol \tau }  \right\rangle  \left(I_{n^2+n} -J_{\mu 2} \left\langle {\rm e}^{  A_{\overline{ \mu }}  \boldsymbol T_s } \right \rangle  \right)^{-1}  \times \\ &
\left( R_{\mu 2}  +  J_{\mu 2}  \left \langle  {\rm e}^{ A_{\overline{\mu }} \boldsymbol T_s } \int_0^{\boldsymbol T_s } {\rm e}^{- A_{\overline{\mu}} r} \hat{a}_{\overline{ \mu }} dr  \right \rangle \right)\\ & + \left\langle     {\rm e}^{  A_{\overline{ \mu }}  \boldsymbol \tau }  
 \int_0^{\boldsymbol \tau}  {\rm e}^{-  A_{\overline{ \mu }} r}  \hat{a}_{\overline{ \mu }} dr   \right\rangle .
	\end{aligned}
	\end{equation} 
\end{theorem}$\et$\\
The proof of this theorem is omitted due to space constraints.

\setcounter{equation}{36}
\begin{figure*}[!th]
	\vspace{2mm}
	\begin{equation}
	\begin{aligned}
	CV^2  = \frac{\overline{ \langle \bold{x}^2 \rangle}-\overline{ \langle \bold{x} \rangle}^2}{\overline{ \langle \bold{x} \rangle}^2}=&\overbrace{\frac{-8 \left(1 - \frac{1}{4}\langle e^{-2\gamma  \boldsymbol T_s}\rangle  \right)\left(1 -\langle e^{-\gamma  \boldsymbol T_s}\rangle  \right)^2  +4 \gamma \langle \boldsymbol T_s \rangle \left(1 - \frac{1}{4}\langle e^{-\gamma  \boldsymbol T_s}\rangle^2  \right)\left(1 -\langle e^{-2\gamma  \boldsymbol T_s}\rangle  \right)  }{8 \left(1 - \frac{1}{4}\langle e^{-2\gamma  \boldsymbol T_s}\rangle  \right)\left(- 1 +  \langle e^{-\gamma  \boldsymbol T_s}\rangle +2 \gamma \langle \boldsymbol T_s \rangle (1- \frac{1}{2}\langle e^{-\gamma  \boldsymbol T_s}\rangle)\right)^2}}^\text{Fluctuations contributed from noisy cell-cycle times}\\ 
	& + \overbrace{ \frac{1-\frac{1}{2} \langle e^{-\gamma  \boldsymbol T_s}\rangle }{8(1 - \frac{1}{4}\langle e^{-2\gamma  \boldsymbol T_s}\rangle) }  \frac{ -3 +8 \gamma \langle \boldsymbol T_s \rangle -  \langle e^{-2\gamma  \boldsymbol T_s}\rangle(2\gamma \langle \boldsymbol T_s \rangle-3  )}{- 1 +  \langle e^{-\gamma  \boldsymbol T_s}\rangle +2 \gamma \langle \boldsymbol T_s \rangle (1- \frac{1}{2}\langle e^{-\gamma  \boldsymbol T_s}\rangle)}   \frac{U}{ \overline{\langle \bold x \rangle} }}^\text{Fluctuations contributed from random synthesis of protein}\\
	&+  \overbrace{ b   \frac{1- \langle e^{-2\gamma  \boldsymbol T_s}\rangle }{1 - \frac{1}{4}\langle e^{-2\gamma  \boldsymbol T_s}\rangle }  \frac{1 -  \langle e^{-\gamma  \boldsymbol T_s}\rangle}{- 1 +  \langle e^{-\gamma  \boldsymbol T_s}\rangle +2 \gamma \langle \boldsymbol T_s \rangle (1- \frac{1}{2}\langle e^{-\gamma  \boldsymbol T_s}\rangle)}  \frac{1 }{ \overline{\langle \bold x \rangle} }}^\text{Fluctuations contributed from partitioning of protein molecules}.
	\end{aligned} 
	\label{noise}
	\end{equation}
\end{figure*}
\setcounter{equation}{26}

\begin{figure}[!h]
	\centering
	\includegraphics[width=1\columnwidth]{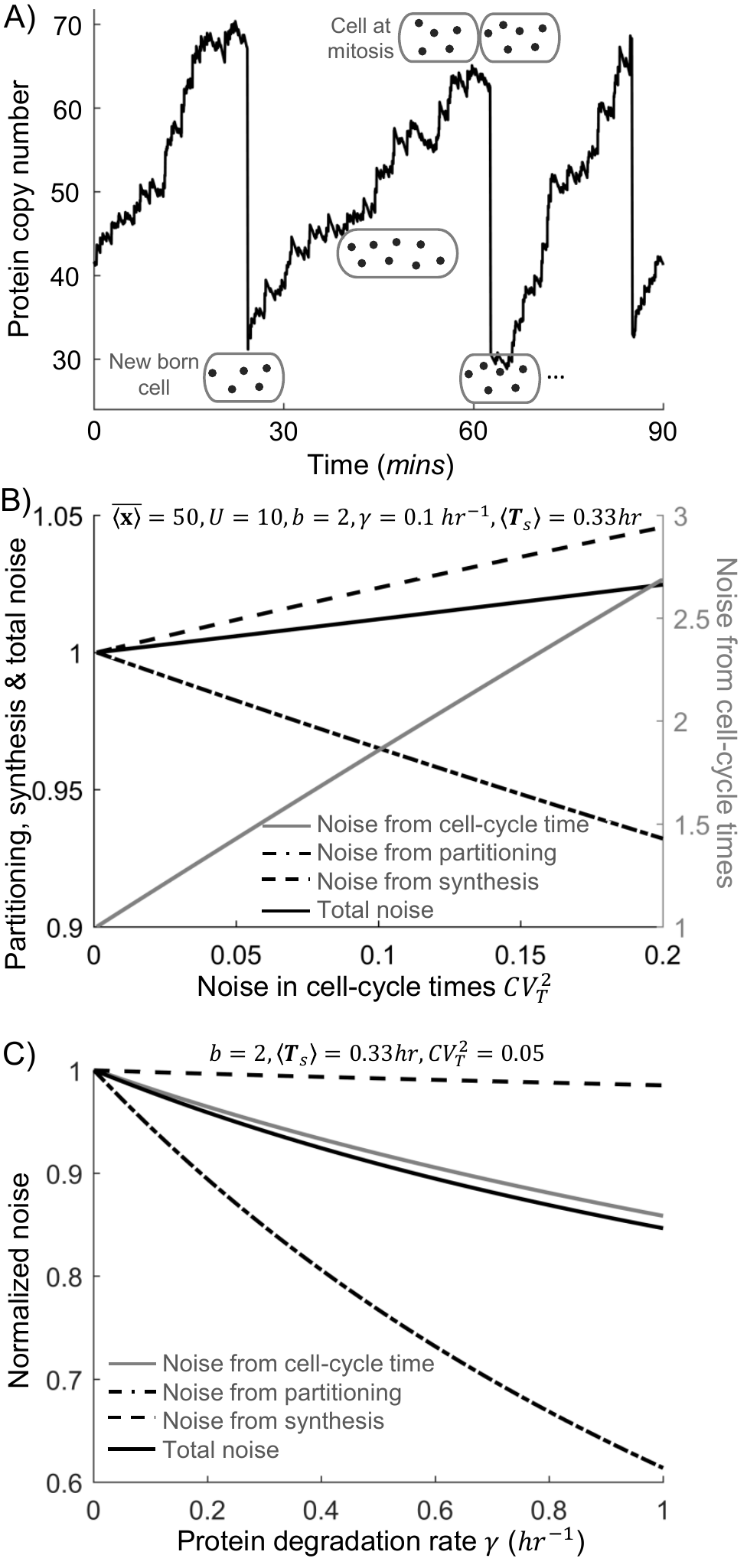}
	\caption{ A) Illustration of a stochastic realization for the protein copy numbers in a single cell. Protein molecules are expressed in random bursts and accumulate within the cell. 	Cell-division events occur at random times, and at the time of division, protein molecules are randomly partitioned between two daughters.  B) Total noise in the level of a protein is decomposed into three components: noise arising from bursty synthesis, random cell-cycle times, and partitioning. While the first two noise components increase with increasing noise in cell-cycle times, the noise contribution from partitioning decreases. C) While noise from bursty synthesis remains almost constant with respect to changes in the protein decay rate $\gamma$, noise contributions from random cell-cycle times and partitioning decrease with increasing $\gamma$. All noise levels are normalized to their value at $CV^2_T=0$ in part B and $\gamma=0$ in part C. For these plots cell-cycle times $\langle \boldsymbol T_s \rangle$ are assumed to be gamma distributed and noise in the cell-cycle times is quantified by the coefficient of variation squared $CV^2_T =\langle \boldsymbol T_s^2 \rangle / \langle \boldsymbol T_s \rangle^2 -1.$}
	\label{fig2}
\end{figure}

\section{Modeling fluctuations in protein copy numbers}
Advances in experimental technologies over the last decade have shown that the level of a given protein in an individual cell is a stochastic process \cite{bpm06,rao08,nmt13,hfh12,brb14,mal13}. These fluctuations in protein levels have been on one hand implicated in corrupting functioning of genetic circuits \cite{lps07,fhg04,leh08}, but on the other hand can be beneficial in terms of creating cell-to-cell phenotypic heterogeneity that is useful to the cell population in changing environments\cite{ele10,vsk08,kul05,bmc04,sdt17,Abranches01072014}.  Diverse noise mechanisms are implicated in driving protein copy number fluctuations, including i) synthesis of proteins
in stochastic bursts \cite{src10,ccg14,smg11,drs12}; ii) random timing of cell-division events that approximately halve the protein count, and iii) randomness in partitioning of protein molecules between two daughters at the time of division \cite{huh11,hup11}. 

Our prior studies in coupling protein synthesis with cell-division events have either assumed a zero protein degradation rate  \cite{sva15,sos16c}, or modeled protein synthesis deterministically  \cite{ans14}. Here we relax these system assumptions, and show that the resulting model falls within the class of PDMPs defined in Section II. Hence, direct application of Theorems 1 \& 2 to the model provide exact results for the protein mean and noise levels with unique biological insights 

\subsection{Model formulation}
Let scalar $\bold x(t)$ denote the protein population count level inside a single cell at time $t$. Assuming Poisson arrival of transcription events, protein production occurs
at exponentially-distributed time intervals with rate $k$. Each synthesis event increases the protein count by  $U$
\begin{equation}
 \bold x \mapsto \boldsymbol x+U, \label{U}
 \end{equation}
 where $U$ can be interpreted as the ``burst size",  and comparing the above reset to \eqref{reset}  $J_1=1$ and $R_1=U $. Further, we assume that protein degradation is deterministic and follows first-order kinetics
 \begin{equation}
 \frac{d\bold{x}}{dt}=- \gamma \bold{x}(t), \label{count}
 \end{equation}
where $\gamma$ is the degradation rate. Our prior work has shown that deterministic decay is a good approximation for large enough protein burst sizes \cite{sbf15,bos15}.
We consider cell-division as a generally-distributed event that changes protein count as
\begin{equation}
\bold{x} \mapsto  \bold x_{2+},
\end{equation}
with the following statistics
\begin{subequations}
	\begin{align}
	& \langle \bold{x}_{2+}  \rangle=\frac{\bold x}{2} , \label{mean+} \\
	&  \langle \bold{x}_{2+}^2 \rangle  - \langle \bold{x}_{2+}\rangle^2 = b \bold{x} \Rightarrow   \langle \bold{x}_{2+}^2 \rangle   =\frac{\bold x^2}{4}+   b \bold{x}, \label{var+} 
	\end{align}	\label{division character}
\end{subequations}
where $b$ quantifies the extent of random partitioning \cite{sva15}. Comparing \eqref{division character} with \eqref{conditional x20}, one can see that here
$J_2=1/2$, $B_2=b/2$, $C_2=1$, and $R_2=Q_2=D_2=0$. 

\subsection{Statistical moments of the protein count}
We aim to derive the mean and variance of protein count. In the first step we define $\boldsymbol\mu \equiv [ \bold x \  \bold x^2 ]^\top$. Using \eqref{count}, in between the events, dynamics of $\boldsymbol \mu$ follows \eqref{mu dynamics} with 
\begin{align} \label{matrix1}
\hat{a}_\mu =0, \ A_\mu =  \left[\begin{array}{cc}
-\gamma & 0\\ 0 & -2 \gamma
\end{array}\right].
\end{align}
Moreover, observing \eqref{U}, the states of the system after exponentially distributed synthesis events change according to \eqref{reset mu} with matrices 
\begin{align}\small
J_{\mu1}= \left[\begin{array}{ccc}
1 & 0 \\ 2k U    & 1
\end{array}\right]  , \ R_{\mu1}=  \left[\begin{array}{c}
k U \\ k  U^2    
\end{array}\right] . 
\end{align}
Hence the matrices $A_{\overline{\mu}}$ and $\hat{a}_{\overline{\mu}}$ in \eqref{mu bar} can be derived as 
\begin{align}\small
\hat{a}_{\overline{\mu}}=\left[\begin{array}{c}
k U  \\ k  U^2  
\end{array}\right] ,  A_{\overline{\mu}} =\left[\begin{array}{ccc}
-\gamma & 0 \\ 2k U    &-2\gamma 
\end{array}\right] .
\end{align}
Further after division, protein count level follows \eqref{reset 2} with matrices
\begin{align} \label{matrix2}
J_{\mu2} = \left[\begin{array}{cc}
1/2 & 0\\ 0  & 1/4
\end{array}\right] , \ R_{\mu2} =0.
\end{align}

With this setup we can use Theorem 2 to calculate the moments of protein count up to order two. In Appendix C we derived all the matrices and vectors needed to calculate steady-state moments. Using them, the mean of protein count can be written as 
\begin{equation}
\overline{ \langle \bold{x} \rangle}=   \frac{k U}{\gamma } -  \frac{k U }{2\gamma^2 \langle \boldsymbol T_s \rangle }  \frac{1 -  \langle e^{-\gamma  \boldsymbol T_s}\rangle }{1 -  \frac{1}{2}\langle e^{-\gamma  \boldsymbol T_s}\rangle } , \label{unstable protein}
\end{equation}
where $\langle \boldsymbol T_s \rangle $ is the mean cell-cycle time.  
Furthermore, for a protein whose half-life is considerably longer than the mean cell-cycle time ($\gamma \to 0$), the above formula simplifies to
\begin{equation}
\overline{ \langle \bold{x} \rangle}=  \frac{k U \langle \boldsymbol T_s \rangle  \left(3+CV^2_T\right)}{2}\label{stable protein}.
\end{equation}
Interestingly, equations \eqref{unstable protein} and \eqref{stable protein} show that noisy cell-cycle times increase the average number of proteins. 

\setcounter{equation}{37}

Moreover by having the second-order moment from Theorem 2, we quantify the noise in protein counts using the coefficient of variation ($CV$) squared as shown in \eqref{noise}. Note that this expression is divided into three components, that represent noise contributions from noise in cell-division times,
 bursty synthesis, and random partitioning of protein molecules during cell division. Interestingly, while the first two terms increase with increasing noise in cell-cycle times, the noise contribution from molecular partitioning has an opposite profile  (Fig. 2B). 
 
We further investigate how these different noise components vary with the protein degradation rate. While noise arising from random partitioning and cell-cycle times decrease as the protein stability decreased, noise from bursty protein synthesis remains relatively constant (Fig. 2C).  In the limit of large degradation rate,  the total noise in protein levels reduces to 
\begin{equation}
\lim_{\gamma\to\infty} CV^2= \text{Noise from synthesis}=\frac{1}{2} \frac{U}{ \overline{\langle \bold x \rangle} } ,
\end{equation}
and is only affected by the randomness in the synthesis process. Moreover, in the limit of stable protein ($\gamma\approx 0$) 
\begin{equation}
\begin{aligned}
CV^2=\overbrace{ \frac{1}{27}+\frac{4\left(9\frac{\langle \boldsymbol T_s^3\rangle}{\langle \boldsymbol T_s  \rangle^3}-9-6CV^2_
	{T}-7CV^4_{T}\right)}{27\left(3+CV^2_{T}\right)^2}}^\text{Noise from cell-cycle times} \\ +\overbrace{\frac{16 b}{3(3+CV^2_{T})}\frac{1}{\overline{ \langle \bold x \rangle}}}^\text{Noise from partitioning}+\overbrace{\frac{3CV^2_{T}+5}{3(3+CV^2_{T})}\frac{U}{\overline{ \langle \bold x \rangle}}}^\text{Noise from synthesis},
\end{aligned}
\end{equation}
where $\langle \boldsymbol T_s^3 \rangle$ denotes the third-order moment of cell-cycle time. This equation illustrates a key point -  the noise in a stable protein just depends on the first three moments of the cell-cycle time. Note that even for deterministic cell-cycle times, deterministic synthesis of molecules, and zero partitioning errors ($b=0$), noise is not zero, i.e. $CV^2=1/27$, and this residual term represents fluctuations in protein levels (i.e linear increase with time and then division by half) within a cell-cycle.

\section{Conclusion}
We have studied statistical moments for a class of PDMPs with two families of resets, allowing the second family to occur at generally-distributed time intervals. Exact solutions of the first- and second-order moments were derived, and applied to the biological problem of stochastic gene expression. Our analysis reports for the first time, formulas for the mean and the noise in the level of a protein in terms of underlying parameters and random processes, leading to important insights illustrated in Fig. 2.
This is straightforward to expand these results to a system with multiple families of resets, each having Poisson arrivals. However, having more than one family of generally-distributed resets is convoluted and will be the subject of future investigation. Finally, the continuous dynamics here are governed via a set of linear time-invariant ODEs. Our preliminary results show that some classes of time-varying differential equations combined with exponentially distributed events have closed moments. Combining these systems with generally-distributed families of resets is another direction of future research.

\section*{Acknowledgment}
\thanks{AS is supported by the National Science Foundation Grant ECCS-1711548.}

\section*{Appendix}
\subsection{Probability density function of timer}
Using forward Kolmogorov equation, the probability distribution of timer $p(\tau)$ at steady state is described through 
\begin{equation}
\frac{\partial p(\tau)}{\partial \tau} = - h(\tau)p(\tau), \ \ \tau>0, \label{cme tau}
\end{equation}
where $\tau$ is the dummy variable for $\boldsymbol{\tau}$.
We start our analysis by taking integral from both sides of \eqref{cme tau}
\begin{equation}
\frac{\partial p(\tau)}{\partial \tau } = h(\tau)p(\tau)\Rightarrow p(\tau) = p_0 {\rm e}^{-\int_0^{\tau} h(y) d y}, \label{this int}
\end{equation}
where $p_0$ is a normalization constants. It can be shown that $p_0=1/\langle \boldsymbol T_s\rangle$ hence $p(\tau)$ can be written as
\begin{equation}
p(\tau) = \frac{1}{\langle \boldsymbol T_s \rangle} {\rm e}^{-\int_0^{\tau} h(y) d y}. \label{prob. tau}
\end{equation}
Note that the probability density function of $\boldsymbol \tau$ is related to probability density function of $\boldsymbol T_s$, but they are not equal. In fact from \eqref{hr} we have 
\begin{equation}
 f(\tau) = h(\tau )  {\rm e}^{-\int_0^{\tau} h(y) d y}.
\end{equation}

\subsection{Proof of Theorem 1}
We prove Theorem 1 in the following steps: we use forward Kolmogorov equation to calculate the joint probability distribution of timer $\boldsymbol\tau$ and the vector of the states $\bold x$. We use this derivation to calculate the steady-state conditional mean $\overline{\langle \bold x\vert \boldsymbol\tau \rangle}$. Finally we un-condition $\overline{\langle \bold x \vert \boldsymbol\tau \rangle}$ to obtain $\overline{\langle \bold x \rangle}$.

Based on forward Kolmogorov equation,  we have the following for joint probability distribution of timer and the states $p(\tau, x)$ in steady state
\begin{equation}\begin{aligned}
& \frac{\partial p(\tau, {x})}{\partial \tau} + \frac{\partial }{\partial {x} }\left( (\hat{a} + A x )  p(\tau,{x})\right)  =\\ & 
 + h_1  p(\tau, J_1^{-1}(x-R_1))-h_1 p(\tau, {x})  -h_2 (\tau) p(\tau, {x}) , \ \ \tau>0,
\end{aligned}
\label{cme}
\end{equation}  
where $x$ is the vector of dummy variables for $\bold x$. Further the operator $\frac{\partial}{\partial x}$ is defined as
\begin{equation}
\frac{\partial}{\partial {x}} = \left[ \frac{\partial}{\partial x_1}, \frac{\partial}{\partial x_2},\ldots,\frac{\partial}{\partial x_n}\right], 
\end{equation}
which is a vector of partial derivative operators.

By having the joint probability distribution, we define conditional mean $\overline{\langle {\bold{x}} \vert \boldsymbol\tau  \rangle}$ as
\begin{equation}
\begin{aligned}
\overline{\langle {\bold{x}} \vert \boldsymbol\tau  \rangle}\equiv\overline{\langle {\bold{x}} \vert \boldsymbol\tau =\tau \rangle}=&\frac{1}{p(\tau )}\int_{0}^{+\infty} x p(\tau, x)dx.
\label{cond. x'}
\end{aligned}
\end{equation}
Taking derivative with respect to $\tau$ from \eqref{cond. x'} results in
\begin{equation}
\begin{aligned}
\frac{\partial \overline{\langle {\bold{x}} \vert \boldsymbol\tau  \rangle }}{\partial \tau } =& -\frac{\frac{\partial p (\tau)}{\partial \tau}}{p^2(\tau )} \int_{0}^{+\infty}x  p(\tau, x)dx\\
&+\frac{1}{p(\tau )}\int_{0}^{+\infty} x\frac{\partial p(\tau, x)}{\partial \tau} 
dx .
\label{derivate00}
\end{aligned}
\end{equation}

In order to calculate $\frac{\partial \overline{\langle {\boldsymbol{x}} \vert \boldsymbol\tau  \rangle }}{\partial \tau } $ we need the expression of $\frac{\partial p(\tau, x)}{\partial \tau}$ and $\frac{\partial p(\tau)}{\partial \tau}$.
Substituting these expressions from \eqref{cme} and \eqref{cme tau} in \eqref{derivate00} and after some algebraic steps we have
\begin{equation}\begin{aligned}
\frac{\partial  \overline{\langle \bold{x} \vert \boldsymbol\tau \rangle} }{\partial \tau} =(\hat{a}+R_1) + (A+h_1 (J_1-I_n)) \langle \bold{x}\vert \boldsymbol\tau\rangle.
\end{aligned} \label{mean dynamics}
\end{equation}
Thus the conditional mean can be derived as
\begin{equation}\begin{aligned}
& \overline{\langle \bold{x} \vert \boldsymbol\tau =\tau  \rangle} = 
 {\rm e}^{ (A+h_1 (J_1 -I_n)) \tau }  \overline{\langle \bold{x} \vert \boldsymbol\tau =0 \rangle}
\\ & + {\rm e}^{ (A+h_1 (J_1-I_n)) \tau } \int_0^\tau  {\rm e}^{- (A+h_1 (J_1 -I_n)) r}(\hat{a}+R_1) dr .
\end{aligned}
\label{conditional x'}
\end{equation} 
In order to calculate $\overline{\langle \bold{x} \vert \boldsymbol\tau=0 \rangle}$ we use equation \eqref{conditional x0} in the main text. Note that in the time of a reset $\boldsymbol\tau \mapsto 0$, hence in the time of a reset $\bold{x} \vert \boldsymbol\tau=T_s \mapsto \bold{x} \vert \boldsymbol\tau=0$. It can be seen that
\begin{equation}\small
\begin{aligned}
&\overline{\langle \bold{x} \vert \boldsymbol\tau=0 \rangle } = \left(I_n -J_2 \left\langle  {\rm e}^{ (A+h_1 (J_1-I_n))  \boldsymbol T_s } \right \rangle  \right)^{-1} \left(R_2  \right. \\ &
\left. J_2  \left \langle  {\rm e}^{(A+h_1 (J_1-I_n)) \boldsymbol T_s } \int_0^{\boldsymbol T_s } {\rm e}^{-(A+h_1 (J_1-I_n)) r} (\hat{a}+R_1) dr  \right \rangle   \right). \label{initial x}
\end{aligned} 
\end{equation} 
using equation \eqref{prob. tau} to uncondition \eqref{conditional x'} with respect to $\boldsymbol \tau$ results in \eqref{mean level01} in the main text. 

\subsection{Deriving the mean and noise in protein count levels}
Using the matrices introduced in \eqref{matrix1}-\eqref{matrix2}, we built different terms in \eqref{mean level}
\begin{equation} 
\begin{aligned}
&   \langle {\rm e}^{A_{\overline{\mu }} \boldsymbol T_s } \rangle  = \left[\begin{array}{cc}
\langle e^{-\gamma \boldsymbol T_s}\rangle & 0\\ 2\frac{k}{\gamma}\left(\langle e^{-\gamma \boldsymbol T_s}\rangle + \langle e^{-2\gamma \boldsymbol T_s}\rangle \right) &   \langle e^{-2\gamma \boldsymbol T_s }\rangle 
\end{array}\right]  ,\\
& \left \langle  {\rm e}^{ A_{\overline{\mu }} \boldsymbol T_s } \int_0^{\boldsymbol T_s } {\rm e}^{- A_{\overline{\mu}} r} \hat{a}_{\overline{ \mu }} dr  \right \rangle   = \\
& \hspace{18mm}
\left[\begin{array}{c}
\frac{k}{\gamma } (1-\langle e^{-\gamma \boldsymbol T_s}\rangle) \\ \\ \begin{array}{c} \frac{k^2  \langle U \rangle^2 }{\gamma^2 } (\left\langle {\rm e}^{ -2\gamma \boldsymbol T_s}  \right \rangle -2 \left\langle {\rm e}^{ -\gamma \boldsymbol T_s }  \right \rangle +1) \\ + \frac{k \langle U^2 \rangle }{2 \gamma } (1- \left\langle {\rm e}^{ -2\gamma \boldsymbol T_s}  \right \rangle  )
\end{array}\end{array}\right],\label{matrices1}
\end{aligned} 
\end{equation}
and 
\begin{equation}
\begin{aligned}
&   \langle {\rm e}^{A_\mu\boldsymbol \tau} \rangle  = \left[\begin{array}{cc}
\langle e^{-\gamma \boldsymbol \tau}\rangle & 0\\ 2\frac{k}{\gamma}\left(\langle e^{-\gamma \boldsymbol \tau}\rangle + \langle e^{-2\gamma \boldsymbol \tau}\rangle \right) &   \langle e^{-2\gamma \boldsymbol \tau}\rangle 
\end{array}\right]  ,\\
&\left \langle {\rm e}^{ A_\mu\boldsymbol \tau  } \int_0^{\boldsymbol \tau}   {\rm e}^{ -A_\mu r} \hat{a}_\mu dr \right \rangle = \\  & \hspace{20mm}
\left[\begin{array}{c}
\frac{k}{\gamma } (1-\langle e^{-\gamma \boldsymbol \tau}\rangle) \\ \\ \begin{array}{c} \frac{k^2 }{\gamma^2 } (\left\langle {\rm e}^{ -2\gamma \boldsymbol \tau}  \right \rangle -2 \left\langle {\rm e}^{ -\gamma \boldsymbol  \tau }  \right \rangle +1) \\ + \frac{k \langle U^2 \rangle }{2 \gamma } (1- \left\langle {\rm e}^{ -2\gamma \boldsymbol \tau }  \right \rangle  )
\end{array}\end{array} \right] .\label{matrices2}
\end{aligned} 
\end{equation}

In the last step we express $\langle {\rm e}^{-\gamma  \boldsymbol \tau} \rangle$ as a function of $\langle {\rm e}^{-\gamma \boldsymbol  T_s } \rangle$
\begin{equation} 
\begin{aligned}
\left\langle {\rm e}^{ A\boldsymbol \tau} \right \rangle&  = \frac{1}{\langle \boldsymbol T_s \rangle }\int_0^{\infty}   {\rm e}^{-\int_0^{\tau} h(y) d y} {\rm e}^{-\gamma \tau }d\tau 
\end{aligned}
\end{equation} 
\begin{equation} \nonumber 
\begin{aligned}
&  = -\frac{1}{\gamma } \frac{1}{\langle \boldsymbol T_s \rangle }\left(  {\rm e}^{-\int_0^{\tau} h(y) d y} {\rm e}^{-\gamma \tau } \right)_0^\infty \\  & - \frac{1}{\gamma }  \frac{1}{\langle \boldsymbol T_s \rangle }\int_0^{\infty}  h(\tau)  {\rm e}^{-\int_0^{\tau} h(y) d y}{\rm e}^{-\gamma \tau }  d\tau \\
&  =\frac{1}{\gamma}  \frac{1}{\langle \boldsymbol T_s \rangle }\left(  I_n - \left\langle {\rm e}^{-\gamma \boldsymbol  T_s}  \right \rangle \right).
\end{aligned}
\end{equation}
Similarly 
\begin{align}\label{T is}
\langle {\rm e}^{-2\gamma\boldsymbol \tau} \rangle  =\frac{1}{\langle \boldsymbol T_s \rangle }\frac{1}{2\gamma}\left(  1- \left\langle {\rm e}^{ -2\gamma \boldsymbol  T_s}  \right \rangle \right). 
\end{align}
Putting everything back to the extension of equation \eqref{mean level}, we can calculate mean and the second-order moment of protein count.

\bibliographystyle{plos2009}
\bibliography{RefMaster}


\end{document}